\documentstyle[twoside,epsf]{article}

\begin{document}
\

\begin{center}

{\LARGE{\bf{THE MANIPULATION PROBLEM}}}

\medskip

{\LARGE{\bf{IN QUANTUM MECHANICS}}\footnote{Talk given at the
workshop `Symmetries in quantum mechanics and quantum optics', Burgos
(Spain), September 1998. To be published in the Proceedings.}}

\vskip1cm

{\sc David J. Fern\'andez C.$^\dagger$}
\vskip0.5cm

{\it $^\dagger$ Depto. de F\'{\i}sica, CINVESTAV-IPN \\ A.P. 14-740, 07000
M\'exico D.F., Mexico}
\end{center}
\bigskip

\begin{abstract}
We explain the meaning of dynamical manipulation, and we illustrate its
mechanism by using a system composed of a charged particle in a Penning
trap. It is shown that by means of appropriate electric shocks (delta-like
pulses) applied to the trap walls one can induce the squeezing
transformation. The geometric phases associated to some cyclic evolutions,
induced either by the standard fields of the Penning trap or by the
superposition of these plus a rotating magnetic field, are analysed. 
\end{abstract}


\section{Introduction}

The quantum control, or wavepacket engeenering, is one of suggestive
subjects in Quantum Mechanics (QM). The very name reflects the dreams of
almost any physicist: to be able to induce a physical system to make
anything we wish \cite{la}-\cite{fr}. Since long time ago we have been
involved with that subject using a specific name: {\it dynamical
manipulation problem}. My goals in this work are diverse: 

\begin{itemize}

\item{} To explain what we mean precisely by dynamical manipulation

\item{} To illustrate how it works in a realistic physical arrangement,
namely the Penning trap

\item{} To show the relation between dynamical manipulation and some `hot'
subjects in QM as squeezing and geometric phases

\end{itemize}
Having this in mind, this work has been organized as follows.  In section
2 some generalities of the manipulation problem will be presented.  Some
specific dynamical processes, {\it the evolution loops}, will be
introduced, and we will explain that they can be used in order to generate
an arbitrary unitary transformation. In particular, they will be useful to
induce the free evolution going back in time. In section 3 those general
techniques will be applied to a realistic arrangement, namely, the Penning
trap. Thus, it will be shown that the evolution loops occur in the Penning
trap, and they will be called Penning loops. The `perturbed' Penning loops
will be used as a starting point to induce the squeezing transformation
(in general the scale operation) and the Fourier-like transformation. In
section 4 the general setting of the geometric (Aharonov-Anandan) phase
will be shortly presented. Section 5 will stablish the connection between
the geometric phase and the Penning loops (perturbed and unperturbed). The
paper will end with some general conclusions. 

\section{Manipulation problem}

\noindent Motivation: at first sight there is an asymmetry in nature so
that the evolution {\it forward} in time is privileged. One might think
that certain unitary operators cannot be dynamically achieved, e.g., the
inverted free evolution towards {\it the past} (backwards in time)
$$
e^{i\lambda {p^2\over 2}}, \qquad \lambda>0.
$$
The dynamical manipulation problem tries precisely to answer the question: 
can any given unitary transformation $U$ on a physical system be
dynamically induced? In other words, can we find a Hamiltonian $H(t)$ such
that $U$ arises from a solution to the Schr\"odinger equation with such a
$H(t)$ at some time $t=\tau$?
$$
U=U(t=\tau), \qquad {d\over dt} U(t) = -i H(t) U(t), \qquad U(t=0) = I.
$$
The essence of the idea was formulated by Lamb in terms of system states,
namely, given any two state vectors $\vert a\rangle$ and $\vert b\rangle$
one looks for a Hamiltonian $H(t)$ which links them dynamically so that
there is a solution $\vert\psi(t)\rangle$ to Schr\"odinger equation
becoming $\vert a\rangle$ and $\vert b\rangle$ at two different times
$t_a, \ t_b$ \cite{la}. Later on, this idea was pursued at the operator
level (as posed above) by Lubkin, Mielnik, Waniewski, the present author
and other colleagues \cite{lu}-\cite{fr}.

There are signs that theoretically any unitary operator can be dynamically
induced if two assumptions are made \cite{mi77}: 

\begin{itemize}

\item{} Any potential $V(q,t)$ continuous in $(q,t)$ and leading to a
self-adjoint Hamiltonian
$$
H(t) = {p^2\over 2} + V(q,t)
$$
can in principle be created; in such a case it is said that
$$
U(t) = {\cal T}\left\{\exp\left[-i\int_{0}^t\left({p^2\over 2}
+ V(q,\tau)\right)d\tau\right]\right\} \eqno(1)             
$$
where ${\cal T}$ is the time ordering symbol, are {\it dynamically
achievable operators} (DAO) \item{} The limit of operators of the form (1)
are DAO

\end{itemize}

\smallskip

\noindent Some simple examples of DAO are:

\smallskip

\begin{itemize}

\item{} $\exp(-it{p^2\over 2}), \ t\geq 0$: induced by the free particle
Hamiltonian $H={p^2\over 2}$

\smallskip

\item{} $\exp[-iV(q)]$: induced by the Hamiltonian associated to the
`kick' of potential $V(q,t) = \delta(t-t_0)V(q)$; it can be seen as the
limit when $\epsilon\rightarrow 0$ of the operator sequence
$$
U_\epsilon=U(t_0,t_0+\epsilon) = e^{-i\epsilon[{p^2\over
2}+{1\over\epsilon}V(q)]} = e^{-i[\epsilon {p^2\over 2} + V(q)]}.
$$

\end{itemize}

\subsection{Evolution loops}

The evolution loops (EL) are specific dynamical processes such that the
evolution operator of the system becomes $I$ (modulo phase) at some time
$\tau$ \cite{mi86,fe92,fr,fe94}: 
$$
U(\tau)=e^{i\phi}I \equiv I.
$$
They are a natural generalization of what happens for the harmonic
oscillator. However, the interest on those processes arose after the
discovery of the following operator identity \cite{mi77}:
$$
\left(e^{-i\lambda {p^2\over 2}}e^{-i{1\over\lambda} {q^2\over 2}}\right)^6
\equiv I, \qquad \lambda> 0.  \eqno{(2)}
$$
Notice that the left hand side of (2) involves just DAO so that the EL are
also DAO. A schematic representation of (2) is shown in figure (1.a),
where the sides of the figure represent intervals of the free evolution of
length $\lambda$ while the vertices represent kicks of potential $q^2/2$
of intensity $1/\lambda$. 
\begin{figure}[htbp]
\centerline{(a) \hskip5cm (b)}
\vspace*{13pt}
\begin{minipage}{4truecm}
\hspace*{1truecm}   
\epsfxsize=4truecm
\epsfbox{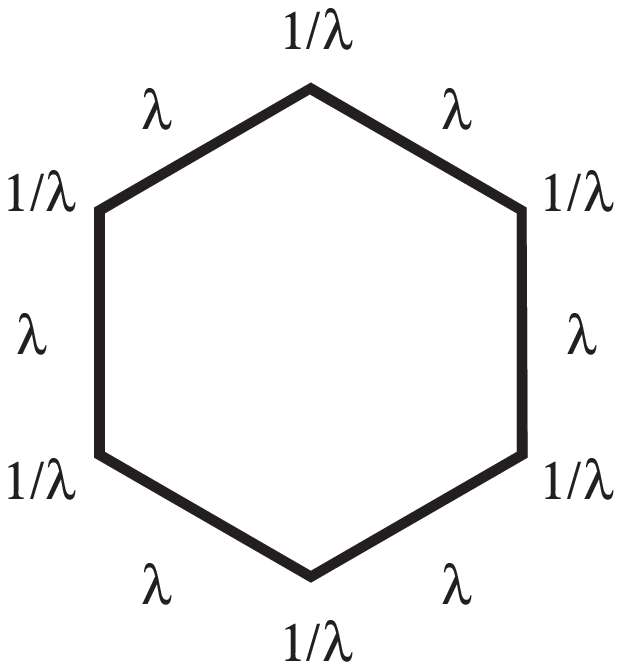}
\hspace*{2truecm}
\epsfxsize=3.3truecm
\epsfbox{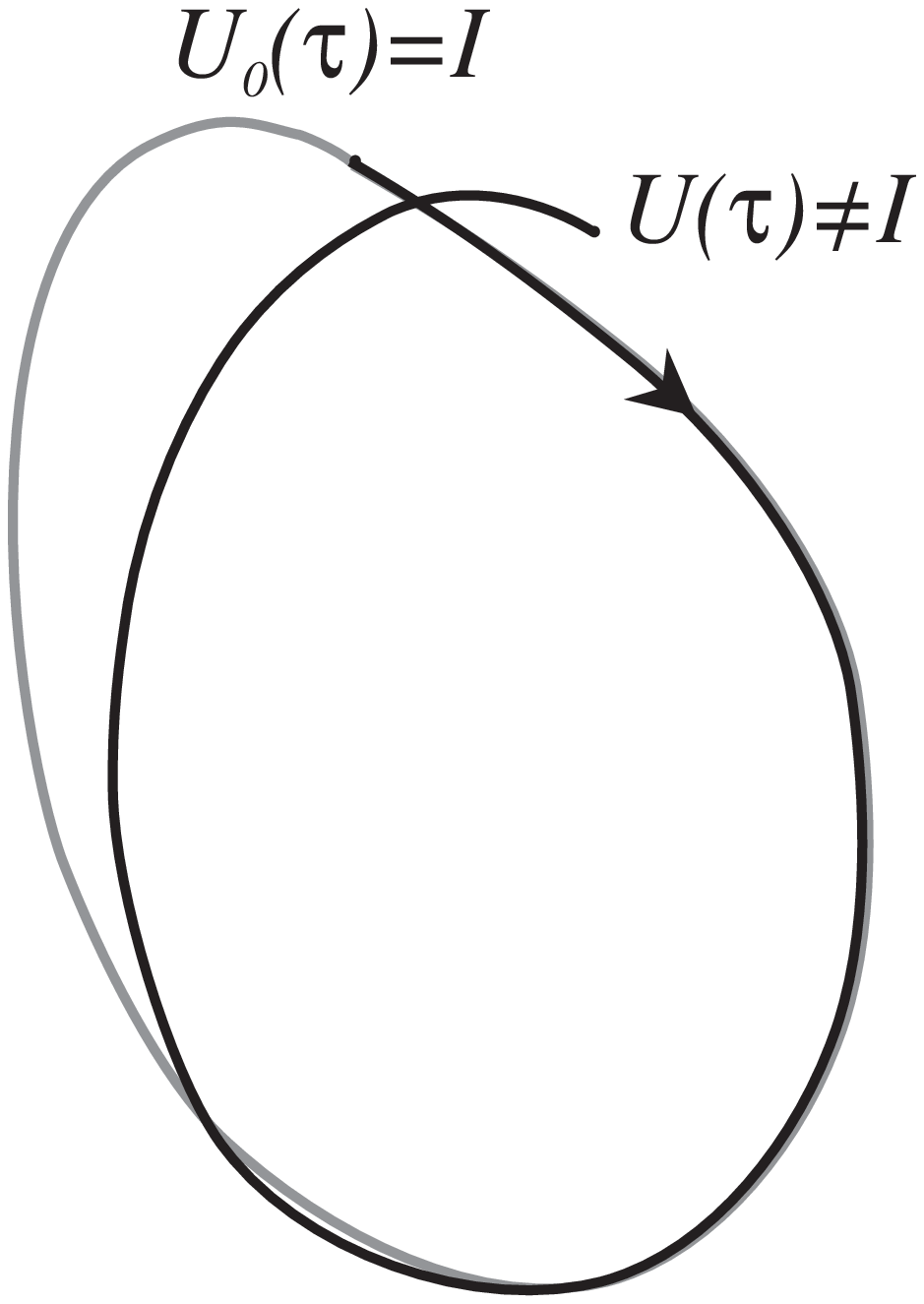}
\end{minipage}
\vspace*{13pt}
\caption{\small a) schematic representation of the evolution loop (2); b) 
the precession of a generic evolution loop induced by some perturbation}
\end{figure}  
It is easy to show (2) in the Heisenberg picture, by considering the LHS
as an evolution operator and noticing that
$$
e^{i{1\over\lambda} {q^2\over 2}} e^{i\lambda {p^2\over 2}}\left(\matrix{q
\cr p}\right) e^{-i\lambda {p^2\over 2}}e^{-i{1\over\lambda} {q^2\over 2}}
= \left(\matrix{1 & \lambda \cr 0 & 1}\right) \left( \matrix{1 & 0 \cr
-{1\over\lambda} & 1}\right) \left(\matrix{q \cr p}\right).
$$

\subsection{Perturbed evolution loops}

The EL are important by the suggestive idea that by applying
perturbations, the complete Hamiltonian (the loop Hamiltonian plus the
perturbation) will induce the precession of the distorted loop which can
lead, in principle, to any unitary operator \cite{mi86}. This process is
schematically represented in figure (1.b), where the gray line represents
the EL and the black line the precession induced by the perturbation.

Some additional interesting applications of the EL can be found.  An
important one arises from equation (2) by noticing that the free evolution
backwards in time is also a DAO
$$
e^{-i{1\over\lambda} {q^2\over 2}}\left(e^{-i\lambda {p^2\over 2}}
e^{-i{1\over\lambda} {q^2\over 2}}\right)^5  \equiv e^{i\lambda {p^2\over
2}}, \qquad \lambda> 0.
\eqno{(3)}
$$
Another suggestive application concerns measurements in QM. To test, e.g.,
the reduction axiom one has to make basically an entire sequence of almost
`simultaneous' measurements (so that the subsequent evolution will not
destroy the wavepacket resulting of the first measurement). For a system
in an EL, once the first measurement has been made we can assure that the
reduced wavepacket, independently of what it was, will be reconstructed at
the finite loop time $\tau$. So, we get more freedom to perform the second
measurement, and the EL is a device avoiding that the reduced wavepacket
will be demolished by the natural evolution \cite{ct}.

\section{Manipulation and Penning trap}

Trying to find a realistic system in which the above techniques could be
applied we arrived to the Penning trap. A charged particle inside an ideal
hyperbolic Penning trap is under the action of a constant homogeneous
magnetic field along $z$-direction plus an electrostatic field induced by
a quadrupolar potential, characterized by the Hamiltonian \cite{fe92,bg}:
$$
H = {1\over 2m}\left({\bf p}-{e\over c}{\bf A}({\bf r})\right)^2 + e
V({\bf r}) ,
$$
$$
{\bf A}({\bf r}) = -{1\over 2} {\bf r}\times{\bf B}_0, \quad V({\bf r})=
V_0({\bf r}^2
-3 z^2).
$$
The region where the charge is trapped is limited by the equipotential
surfaces of $V({\bf r})$. In the real trap, electrodes with that form are
placed at the right positions (see our computer construction in figure 2). 
The two endcaps are at the same potential while the ring is at a different
one. 
\begin{figure}[htbp]
\vspace*{13pt}
\begin{minipage}{11truecm}
\hspace*{1truecm}   
\epsfxsize=11truecm
\epsfbox{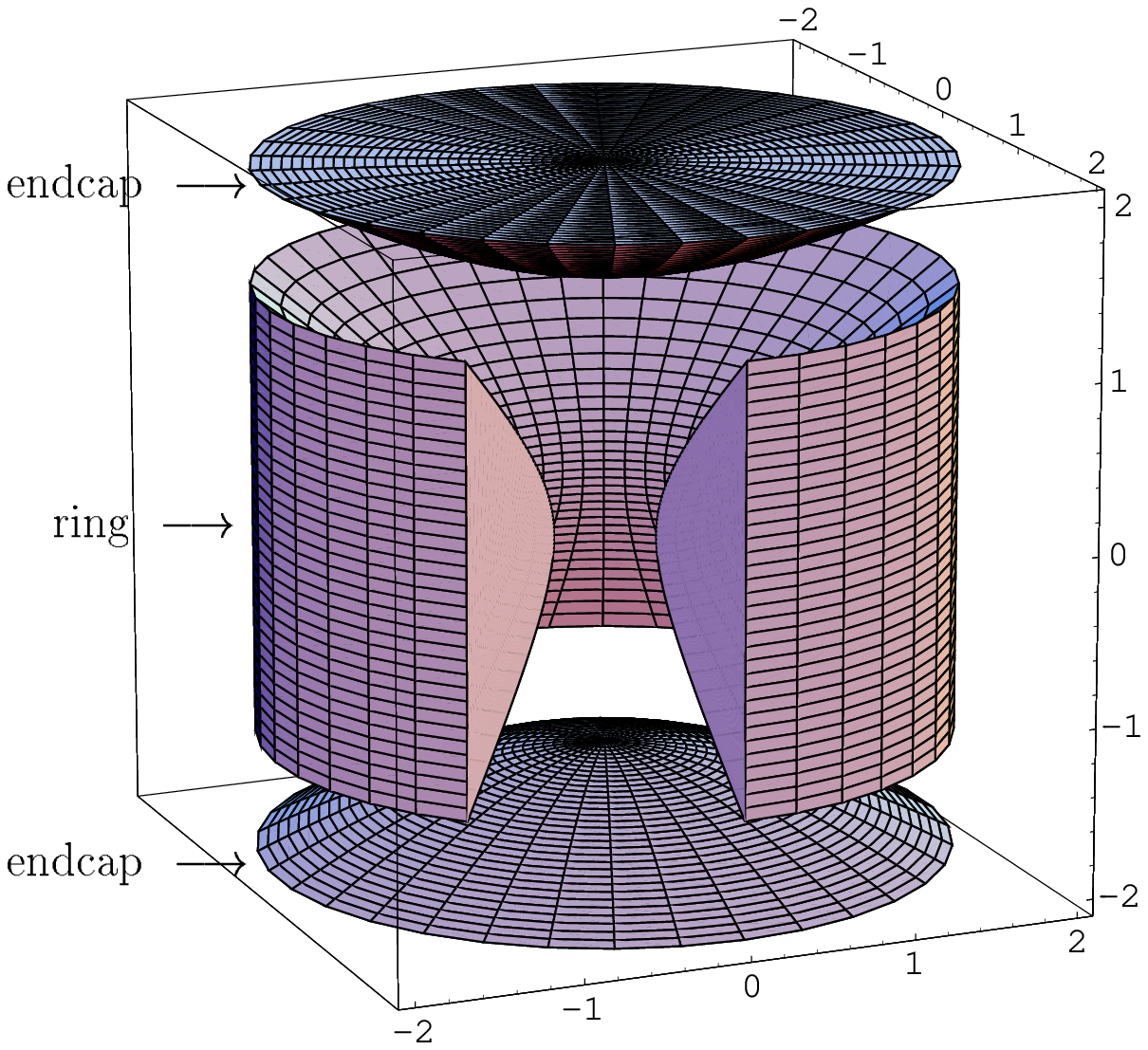}
\end{minipage}
\vspace*{13pt}
\caption{\small simulation of the Penning trap cavity. The missing section
of the ring was removed to see better the inside of the cavity.
}
\end{figure}  

The characteristics of the motion can be easily seen by expanding
explicitly the previous Hamiltonian:
$$
H = H_z + H_c + H_\rho,
$$
where $H_z, \  H_c$ and $H_\rho$ are commuting Hamiltonians given by:
$$
\matrix{H_z = & \hskip-5.2cm {1\over 2m} p_z^2 + {m\over 2} \omega_0^2
z^2, 
\medskip \cr
H_c = &\hskip-6.5cm - {\omega_c\over 2} L_z, \medskip \cr
H_\rho = & {1\over 2m} \left(p_x^2 + p_y^2\right) + {m\over
2}\omega_\rho^2
\, (x^2 + y^2) = {1\over 2m} p_\rho^2 + {m\over 2} \omega_\rho^2 \,
\rho^2,}
$$
and the trap regime is guaranteed if
$$
\omega_0^2 = -{4eV_0\over m} >0, \qquad \omega_c = {eB_0\over mc}>0,
\qquad \omega_\rho^2 = {\omega_c^2 - 2\omega_0^2 \over 4}>0.
$$
Along $z$ direction we have harmonic oscillator motion of frequency
$\omega_0$. On the $x-y$ plane the motion consists of rotations around $z$
axis with frequency $\omega_c/2$ superposed to a harmonic oscillator
motion of frequency $\omega_\rho$ along $\rho$. Notice that there are
indeed just two independent frequencies. 

\subsection{Penning loops}

We want not only trapping, but to induce some finer manipulations on the
system, e.g., the evolution loops. They arise if the three motions $H_z, \
H_c, \ H_\rho$ are `sinchronized', i.e., if the frequencies $\omega_0, \
\omega_c, \ \omega_\rho$ are commensurable. There are different
possibilities:

\begin{itemize}
\item{} If $\omega_\rho/\omega_0 = l_2/l_1 \in {\bf Q}$, the $z-\rho$
motions can be sinchronized. What is left at $\tau = l_1 T, \ T=
2\pi/\omega_0$ are pure rigid rotations

\item{} If $\omega_c/\omega_0 \in {\bf Q}$, it is not guaranteed that
$\omega_\rho/\omega_0$ will be also rational. However, there are some
values of $\omega_c/\omega_0\in{\bf Q}$ for which this happens
$$\matrix{
{\omega_c/\omega_0} = {3/2}, & {\omega_\rho/\omega_0}
={1/4}, & \tau = 2T, \smallskip \cr
{\omega_c/\omega_0} = {9/4}, & {\omega_\rho/\omega_0}
={7/8}, &  \tau = 4T, \smallskip \cr
{\omega_c/\omega_0} = {33/8}, & {\omega_\rho/\omega_0}
= {31/16}, & \tau = 8T.}
$$
\end{itemize}
The results above mean that the evolution loops are DAO of the charge
inside the Penning trap. We have called them Penning loops (PL) 
\cite{fe92}.

\subsection{Perturbed Penning loops}

From now on, let us restrict ourselves to the PL with period $\tau = 2 T$,
i.e., take $\omega_c = 3\omega_0/2, \ \omega_\rho = \omega_0/4$. Let us
`perturb' this PL by a sucession of two instantaneous discharges applied
to the walls of the trap, represented by the potential
$$
V'({\bf r},t) = \left[V_0'\delta(t-t_1) + V_0''\delta(t-t_2)\right] ({\bf
r}^2-3 z^2) \quad t_1<t_2<\tau.
$$
The total Hamiltonian is again a sum of 3 commuting terms
$$
H(t) = H^l + eV'({\bf r},t) = H_z(t) + H_c^l + H_\rho(t),
$$
$$
H_z(t) = H_z^l + {m\over 2}\left[F'\delta(t-t_1) + F''\delta(t-t_2)\right]
z^2,
$$
$$
H_\rho(t) = H_\rho^l - {m\over 4}\left[F'\delta(t-t_1) + 
F''\delta(t-t_2)\right] \rho^2,
$$
$$
F' = - {4eV_0'\over m}, \qquad F'' = - {4eV_0''\over m},
$$
where the superindex $l$ means to choose the specific values of $\omega_0,
\ \omega_c$ and $\omega_\rho$ producing the EL with $\tau = 2T$.  The
evolution operator at $\tau$ takes the factorized form
$$
U(\tau) = U_z(\tau) U_c(\tau) U_\rho(\tau),
$$
where
$$
\matrix{U_z(\tau) = & 
e^{-iH_z^l(\tau-t_2)}e^{-iF''mz^2/2}e^{-iH_z^l(t_2-t_1)}
e^{-iF'mz^2/2}e^{-iH_z^lt_1}, \medskip \cr
U_c(\tau) = & \hskip-5.5cm e^{-iH_c^l\tau} = e^{i3\pi L_z}, \medskip \cr
U_\rho(\tau) = & \hskip-0.5cm
e^{-iH_\rho^l(\tau-t_2)}e^{iF''m\rho^2/4}e^{-iH_\rho^l(t_2-t_1)}
e^{iF'm\rho^2/4}e^{-iH_\rho^lt_1}.}
$$
For quadratic Hamiltonians, $U(\tau)$ is defined by the linear
transformation
$$
U^\dagger(\tau)\left(\matrix{{\bf r} \cr {\bf p}}\right) U(\tau) 
= {\bf u}(\tau)\left(\matrix{{\bf r} \cr {\bf p}}\right),
$$
where ${\bf u}(t)$ is a $6\times 6$ simplectic matrix called {\it
evolution matrix}. In our case that matrix reduces to $2\times2$
unimodular matrices $u_x=u_y={\bf u}_x(\tau)={\bf u}_y(\tau)$, $u_z={\bf
u}_z(\tau)$ acting on the pairs $(x,p_x)^T$, $(y,p_y)^T$, $(z,p_z)^T$:
$$\matrix{
u_x =  & \hskip-0.3cm -u_{ho}({\omega_0\over4},\tau-t_2) 
u_k(-{F''\over2}) u_{ho}({\omega_0\over4},t_2-t_1) 
u_k(-{F'\over2}) u_{ho}({\omega_0\over4},t_1), \medskip \cr 
u_z = &  \hskip-0.8cm u_{ho}(\omega_0,\tau-t_2) u_k(F'')
u_{ho}(\omega_0,t_2-t_1) 
u_k(F') u_{ho}(\omega_0,t_1).} \eqno{(4)}
$$
Notice that the evolution matrices $u_{ho}$, $u_k$, and the corresponding
operators inducing them are of the form: 
$$\matrix{
u_{ho}(\omega,t) = \left(
\matrix{\cos\omega t & {\sin\omega t \over m\omega}\cr
-m\omega \sin\omega t & \cos\omega t}
\right), & U_{ho}(t) = e^{-i({p^2\over 2m} + m\omega^2{q^2\over2})t},
\medskip \cr
\hskip-1.2cm u_k(F)= \left(\matrix{1 & 0 \cr -mF & 1}
\right),
& \hskip-1.1cm U_k(t_0) = e^{-imF{q^2\over2}}.}
$$

We are interested in the following transformations (the corresponding
parameters making them true are immediately reported in the corresponding
table):

\smallskip 

\begin{itemize}

\item{} 3-dim Fourier-like transformation
$$
\matrix{x\rightarrow \lambda_2 p_x & y\rightarrow \lambda_2 p_y &
z\rightarrow \lambda_1 p_z \smallskip \cr
p_x \rightarrow -{1\over\lambda_2} x & p_y \rightarrow -{1\over\lambda_2}
y & p_z \rightarrow -{1\over\lambda_1} z}
$$

\bigskip

\centerline{
\begin{tabular}{|c|c|c|c|c|}   \hline\hline
$\omega_0 t_1$  & $\omega_0 t_2$  & ${F'/\omega_0} =
{F''/\omega_0}$
& $m\omega_0\lambda_1$ &  $m\omega_0\lambda_2$ \\
\hline\hline
$5.3131$  & $\ 7.2533$   & $\ \ 1.5165$  & $\ \ 3.5238$ &  $-39.0332$ \\
\cline{1-5}
$5.3131$  & $\ 7.2533$   & $-1.5165$ & $-0.6046$ &  $\ \ \ \ 6.6970$ \\
\cline{1-5}
$0.9701$  & $11.5962$  & $\ \ 1.5165$  & $\ \ 0.6046$ &   $\ \ -2.3891$ \\
\cline{1-5}
$0.9701$  & $11.5962$  & $-1.5165$ & $-3.5238$ &  $\ \ \ \ 0.4099$ \\
\cline{1-5}
\end{tabular}}

\medskip

\centerline{{\small Table 1. The parameters producing a 3-dim Fourier-like
transformation.}}

\smallskip

\item{} 1-dim Fourier-like transformation in $z$ plus 2-dim scale
transformation in $x-y$
$$
\matrix{x\rightarrow \lambda_2 x & y\rightarrow \lambda_2 y &
z\rightarrow \lambda_1 p_z \smallskip \cr
p_x \rightarrow {1\over\lambda_2} p_x & p_y \rightarrow {1\over\lambda_2}
p_y & p_z \rightarrow -{1\over\lambda_1} z}
$$

\bigskip

\centerline{
\begin{tabular}{|c|c|c|c|c|}   \hline\hline
$\omega_0 t_1$  & $\omega_0 t_2$  & ${F'/\omega_0} =
{F''/\omega_0}$ & $m\omega_0\lambda_1$ &  $\lambda_2$ \\
\hline\hline
$1.2094$  & $\ 5.0738$ & $-2.1381$ & $-6.3874$ &  $-10.2781$ \\
\cline{1-5}
$1.9322$  & $\ 4.3510$ & $-2.1381$ & $-1.0959$ &  $\ \ -3.6353$ \\
\cline{1-5}
$7.4926$  & $11.3569$  & $-2.1381$ & $-6.3874$ &  $\ \ -0.0973$ \\
\cline{1-5}
$8.2153$  & $10.6342$  & $-2.1381$ & $-1.0959$ &  $\ \ -0.2751$ \\
\cline{1-5}
\end{tabular}}

\medskip

{\small Table 2. The conditions to produce 1-dim Fourier-like and 2-dim
scale transformations on $z$ and $x-y$ respectively.}

\smallskip

\item{} 3-dim scale transformation
$$
\matrix{x\rightarrow \lambda_2 x & y\rightarrow \lambda_2 y &
z\rightarrow \lambda_1 z \smallskip \cr
p_x \rightarrow {1\over\lambda_2} p_x & p_y \rightarrow {1\over\lambda_2}
p_y & p_z \rightarrow {1\over\lambda_1} p_z}
$$

\bigskip

\centerline{
\begin{tabular}{|c|c|c|c|c|c|}   \hline\hline
$\omega_0 t_1$  & $\omega_0 t_2$  & ${F'/\omega_0}$ & ${F''/\omega_0}$
& $\lambda_1$ &  $\lambda_2$ \\ \hline\hline
$1.2363$  & $\ 8.4896$   & $-1.1589$  & $\ \ 0.7524$ &  $0.4712$ &
$5.0901$ \\ \cline{1-6}
$2.2064$  & $\ 7.5194$   & $-0.7524$ & $\ \ 1.1589$ &  $2.1222$  &
$5.0901$ \\ \cline{1-6}
$4.0768$  & $11.3301$  & $\ \ 0.7524$  & $-1.1589$ &   $2.1222$
& $0.1965$ \\ \cline{1-6}
$5.0469$  & $10.3600$  & $\ \ 1.1589$ & $-0.7524$ &  $0.4712$ &
$0.1965$ \\ \cline{1-6}
\end{tabular}}

\medskip

{\small Table 3. The parameters inducing the 3-dim scale transformation.}

\end{itemize}

\bigskip

To find the values producing, e.g., the 3-dimensional Fourier-like
transformation, we impose that the off diagonal elements of $u_x$ and
$u_z$ of (4) become null. By solving these equations we will find the
values of $\omega_0 t_1, \ \omega_0 t_2, \ F'/\omega_0$ and $
F''/\omega_0$ inducing that transformation, and from these it is simple to
find the corresponding values of the parameters $\lambda_1$ and
$\lambda_2$ (see table 1). The same procedure can be used for the other
two transformations (see tables 2 and 3) \cite{fe92}. 

We are specially interested in the scale transformation due to its close
connection with squeezing. From table 3, it is clear that we can produce
different combinations of squeezing and amplification. For instance, if we
take the values in the first row, at the end of the full process we will
have produced a squeezing with scaling $0.47$ along $z$ direction and an
expansion on $x-y$ scaled by $5.09$. By taking the values in the fourth
row we will have produced the 3-dimensional squeezing with scalings $0.47$
along $z$ and $0.19$ on $x-y$ plane. It is interesting to notice that a
particular scale transformation can also be gotten if
$$
\omega_0 t_2 = \omega_0 t_1 + 2\pi, \qquad {F''\over\omega_0} =
- {F'\over\omega_0} = \cot({\omega_0 t_1\over2}).
$$
The scaling parameters are in this case (see figure 3):
$$
\lambda_1 = 1,   \qquad \lambda_2 = \left({1+ \cos({\omega_0 t_1\over2}) 
\over \sin({\omega_0 t_1\over2})}\right)^2, \qquad 0<\omega_0 t_1<2\pi.
$$
This means that along $z$ direction an EL is again produced but on the
$x-y$ plane we have gotten the scale transformation. The squeezing arises
when $\omega_0 t_1$ takes values in the interval $(\pi,2\pi)$ with the
corresponding kick intensities as given above.
\begin{figure}[htbp]
\begin{minipage}{7truecm}
\hspace*{1.5truecm}   
\epsfxsize=7truecm
\epsfbox{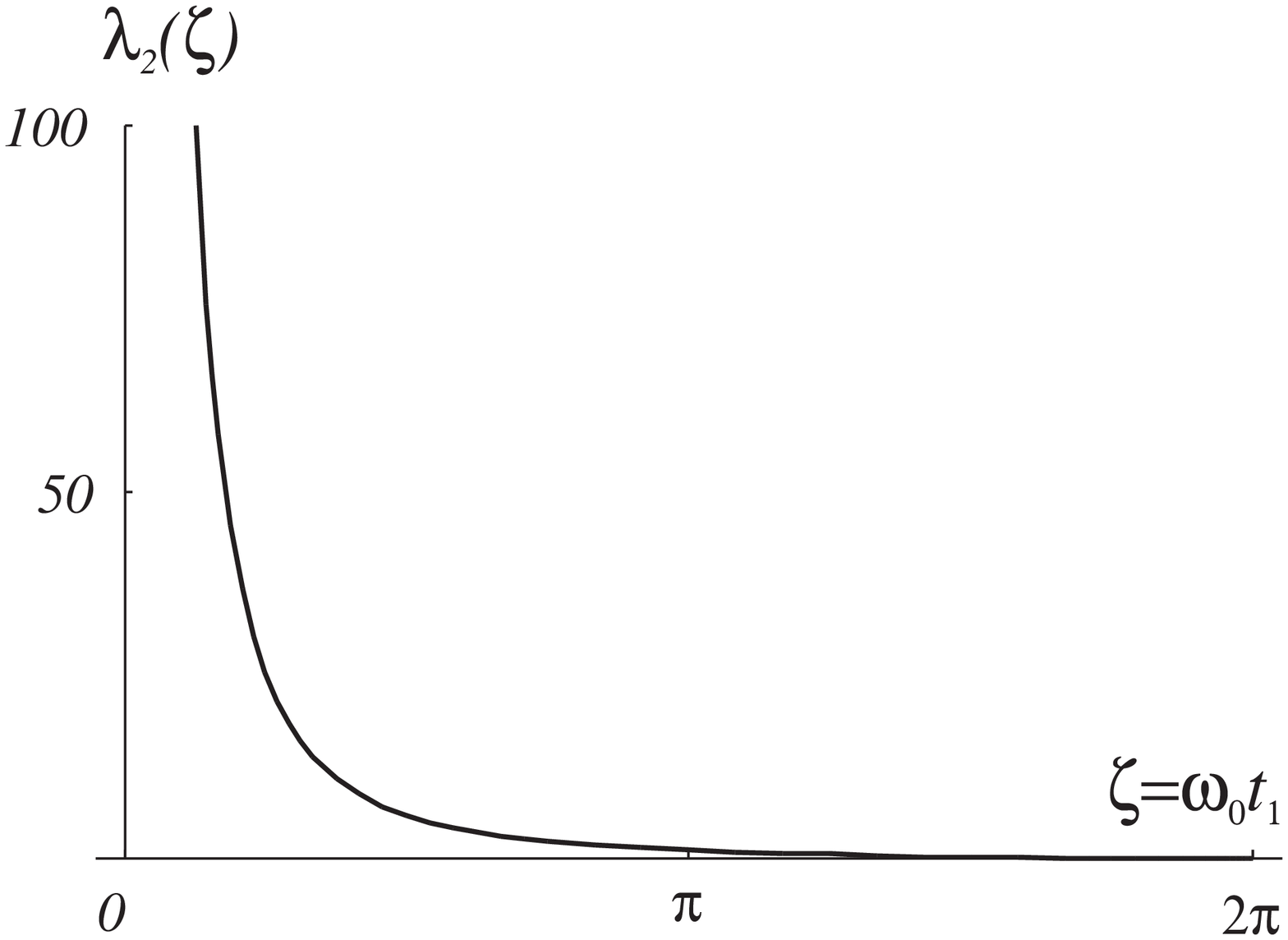}
\end{minipage}
\vspace*{13pt}
\caption{\small the scale parameter $\lambda_2$ as a function of
$\zeta=\omega_0 t_1$.}
\end{figure}  

\section{Geometric phase}

The geometric phase was discovered by Berry, who realized that for cyclic
adiabatic evolution of the eigenstates of a slowly changing cyclic
Hamiltonian, $H(\tau) = H(0)$, there is associated a geometric phase
factor unnoticed by many people that had been working by years with the
adiabatic approximation \cite{be}. Later on, Aharonov and Anandan realized
that the key point of Berry's approach was the cyclicity of the state
rather than the adiabatic assumption, and thus they associated a geometric
phase to any cyclic evolution regardless whether or not the Hamiltonian
inducing the evolution changes slowly in time or even if it is
time-independent \cite{aa}. The most economic formulation of the geometric
phase runs as follows \cite{bbk}. 

Suppose that a system state $\vert\psi(t)\rangle$ is cyclic, i.e.
$$
\vert\psi(\tau)\rangle = e^{i\phi}\vert\psi(0)\rangle.
$$
It turns out that $\phi$ is a sum of a dynamic plus a geometric
contribution, the last one called geometric phase is given by
$$
\beta = \phi + i\int_0^\tau \langle\psi(t)\vert{d\over dt}\vert \psi(t)
\rangle dt.
$$
$\beta$ is geometric in the sense that it is the holonomy of the
horizontal lifting of the closed trajectory in the projective Hilbert
space ${\cal P}$, which arises due to the curvature of ${\cal P}$. Hence,
$\beta$ measures global curvature effects of the projective Hilbert space.
It is worth to notice that the calculation of the geometric phase, even
the determination of the cyclic states of a given system, is not an easy
task. This is one of the reasons why a lot of people have became involved
in the subject \cite{fr,fe94,fnos,fos,fb,cl}.

\section{Geometric phase and Penning loops}

\smallskip

As pointed above, for a system in a Penning loop any state
$\vert\psi(t)\rangle$ is cyclic with period equal to the loop period: 
$$
\vert\psi(\tau)\rangle = e^\phi \vert\psi(0)\rangle.
$$
Thus, it is quite natural that a geometric phase factor should be
associated to any state. For evolution loops induced by time-independent
Hamiltonians, the geometric phase is directly related to the expected
value of the energy in the initial state:
$$
\beta = \phi + \tau \langle\psi(0)\vert H^l\vert\psi(0) \rangle.
$$
An alternative formula arises working in the basis of energy eigenstates
of $H^l$, where again the superindex $l$ means that we are taking the
parameters of $H$ so as to induce the corresponding Penning loop. Let be
$\vert E_{n_1n_2n_3}\rangle$ the eigenstate of $H^l$ associated to the
eigenvalue $E_{n_1n_2n_3}$: 
$$
H^l \vert E_{n_1n_2n_3}\rangle = E_{n_1n_2n_3} \vert E_{n_1n_2n_3}\rangle.
$$
By decomposing now $\vert\psi(0) \rangle$ in that basis
$$
\vert\psi(0) \rangle = \sum_{n_1n_2n_3=0}^\infty c_{n_1n_2n_3} \vert
E_{n_1n_2n_3}\rangle,
$$
we finally get
$$
\beta = \phi + \sum_{n_1n_2n_3=0}^\infty
E_{n_1n_2n_3}\vert c_{n_1n_2n_3}\vert^2.
$$
This formula is similar to the one obtained for the evolution loop induced
by the harmonic oscillator Hamiltonian \cite{fe94} (see also \cite{fr}).

\subsection{Geometric phase and perturbed Penning loops}

\smallskip

Now, instead of `perturbing' the PL by a term affecting the scalar
potential of $H$, as in section 3.2, we perturb its magnetic part so that
the static initial magnetic field ${\bf B}_0$ becomes the rotating
magnetic field \cite{fb}: 
$$ 
{\bf B}(t) = (B\cos\omega t, B\sin\omega t,B_0).  
$$ 
This system is closer to the systems used by other people to analyse the
geometric phase, so it is important to determine its cyclic states. The
Hamiltonian can be written
$$ 
H(t) = {1\over 2m}\left({\bf p} +
{e\over 2c}{\bf r}\times {\bf B}(t)  \right)^2 + {m\over
2}\omega_0^2\left( z^2 - {x^2+y^2\over2} \right).  
$$ 
In order to eliminate the time dependence, let us make the `transition to
the rotating frame' \cite{mf89}, i.e., express the evolution operator as
follows: 
$$
U(t) = e^{-i\omega tL_z}e^{-iGt}, 
$$ 
where $G$ is the time-independent Floquet generator:  
$$ 
G = H(0) - \omega L_z.  
$$ 
As $G$ is quadratic in ${\bf v}^T = ({\bf r},{\bf p})$, the motion is
determined by the kind of linear transformation induced on ${\bf v}$ by
$e^{-iGt}$: 
$$ 
{\bf v}(t) = e^{iGt} {\bf v} e^{-iGt} = e^{\Lambda t} {\bf
v}, \qquad [iG,{\bf v}] = \Lambda {\bf v}.  
$$ 
This depends on the roots of the characteristic polynomial of $\Lambda$,
${\rm Det}(\lambda I-\Lambda)$, which become dependent of three
dimensionless parameters 
$$ 
\alpha = {\vert e\vert B\over 2mc\omega},
\qquad \alpha_0 = {\vert e\vert B_0\over 2mc\omega} = {\omega_c\over
2\omega}, \qquad w = {\omega_0\over \omega}.  
$$ 
Our interest is centered in the case when all the roots of ${\rm
Det}(\lambda I-\Lambda)$ are purely imaginary because in such a case $G$
will induce `confined' motions; this just restricts the parameters
$\alpha, \ \alpha_0, \ w$ to some region in $\alpha-\alpha_0-w$ space. By
simplicity, we assume that the static fields of the Penning trap induce
the PL with period $\tau = 2T$, i.e., 
$$ 
\omega_0 = {2\omega_c\over 3} \quad \Rightarrow \quad w = {2\omega_c\over
3\omega} = {4\alpha_0\over 3}.
$$ 
With this assumption, we can illustrate the classification of the
2-dimensional parameter space $\alpha-\alpha_0$ according to the nature of
the motion induced on the charged particle (see figure 4). The regions for
which the motion is trapped are labelled as $T_i, \ i=1,\dots,4$, while
the rest of the parameter domain produces deconfined motion \cite{fb}.

\begin{figure}[htbp]
\vspace*{13pt}
\begin{minipage}{7truecm}
\hspace*{2.3truecm}   
\epsfxsize=7truecm
\epsfbox{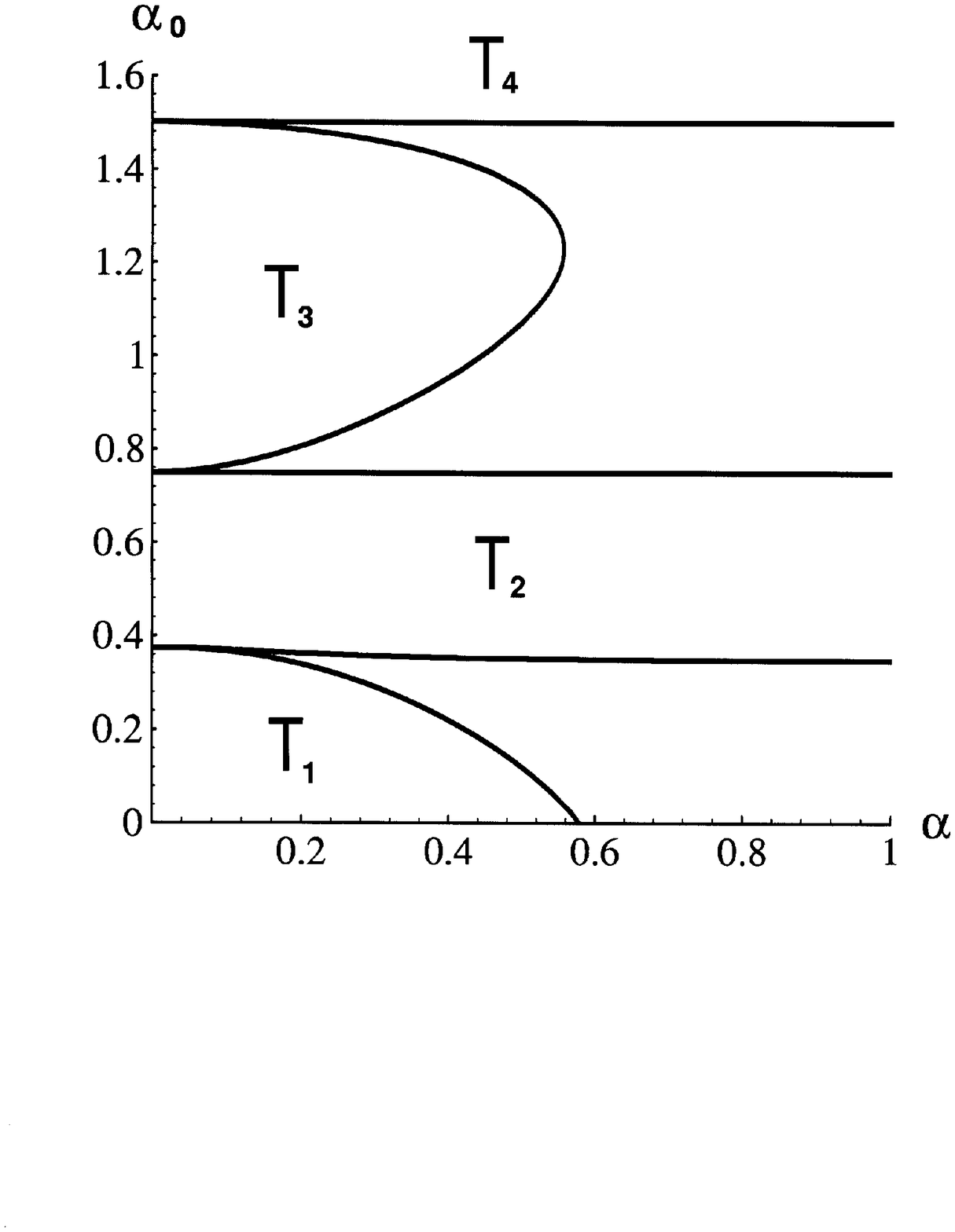}
\end{minipage}
\vspace*{13pt}
\caption{\small classification of the $\alpha-\alpha_0$ plane into regions
in which the charged particle performs confined motion (the regions
labelled by $T_i, \ i=1,\dots,4$) and deconfined motion (the rest of the
plane).} 
\end{figure}  

From now on, let us suppose that the parameters $\alpha$ and $\alpha_0$
belong to one of the regions $T_i, \ i=1,\dots,4$. In such a case, the
Floquet generator $G$ becomes the superposition of three harmonic
oscillator Hamiltonians: 
$$
G = \sum_{i=1}^3 \epsilon_i\omega_i\left(A_i^\dagger A_i +
{\epsilon_i\over 2} \right),
$$
$$
[A_i,A_i^\dagger] = \epsilon_j\delta_{ij}, \qquad \epsilon_i = \pm 1.
$$
Notice that some of the $\epsilon_i, i=1,2,3$ could be negative, and in
such a case there will be a global minus for the oscillator Hamiltonian
contributing to $G$; this will be reflected in the spectrum of $G$ which
will not be bounded by below (see also \cite{mf89}). 

Denote the eigenstates of $G$ by $\vert {\cal E}_{n_1n_2 n_3}\rangle$,
i.e.
$$
G\vert {\cal E}_{n_1n_2n_3} \rangle = {\cal E}_{n_1n_2n_3} \vert {\cal
E}_{n_1n_2n_3} \rangle.
$$
Let us assume that
$$
\vert\psi(0)\rangle = \vert {\cal E}_{n_1n_2n_3} \rangle.
$$
The time evolution of this state is very simple:
$$
\vert\psi(t)\rangle = e^{-i\omega tL_z}e^{-iGt} \vert {\cal E}_{n_1 n_2
n_3} \rangle
= e^{-i{\cal E}_{n_1n_2n_3}t} e^{-i\omega tL_z}\vert {\cal E}_{n_1 n_2
n_3} \rangle.
$$
Notice that this state is cyclic with period $\tau = 2\pi/\omega$
$$
\vert\psi(\tau)\rangle = e^{-i{\cal E}_{n_1n_2n_3}\tau}
\vert\psi(0)\rangle.
$$
Its corresponding geometric phase is easily evaluated
$$
\beta_{n_1n_2n_3} = 2\pi \langle {\cal E}_{n_1n_2n_3} \vert L_3 \vert
{\cal E}_{n_1n_2n_3}\rangle.
$$
It is possible to express the previous phase in terms of ${\cal
E}_{n_1n_2n_3}$; indeed, we have gotten a beautiful expression
\cite{fos,fb}:
$$
\beta_{n_1n_2n_3} = -2\pi{\partial\over\partial\omega} {\cal
E}_{n_1n_2n_3} = -2\pi \sum_{i=1}^3\epsilon_i \left(n_i + {1\over2}\right)
{\partial\omega_i\over \partial\omega}.
$$
Thus, if we change slightly the rotation frequency $\omega$ of the field
there will be a slight change in the levels of the Floquet generator $G$.
The difference of the new level position and the old one at first order in
$\omega$ is essentially the geometric phase. 

\newpage

\section{Conclusions}

\begin{itemize}

\item{} We have seen that the dynamical manipulation, a procedure which
looks for the variety of the operations that can be dynamically induced on
a physical system, provides a more global notion of dynamics than the
standard one.

\item{} We have seen also that the dynamical manipulation leads in a
natural way to the study of some fundamental and practical problems in QM,
as the squeezing transformation and the wavepacket reduction in a
non-demolishing arrangement. 

\item{} Finally, we have seen that those special dynamical processes
called evolution loops provide the tools to easily evaluate the geometric
phases for the corresponding cyclic evolutions or to simplify their
calculation. 

\end{itemize}

{\section*{Acknowledgments}}
\noindent
The author acknowledges the organizers of the workshop `Symmetries in
quantum mechanics and quantum optics' by their kind invitation to give
this talk at the beautiful town of Burgos (Spain). The support of CONACYT
under project 26329-E is also acknowledged.

\end{document}